\documentstyle[12pt,psfig]{article}
\def\be{\begin{equation}}
\def\ee{\end{equation}}

\def\TL{\hfil$\displaystyle{##}$}

\def\TT{\hbox{##}}
\def\seqalign#1#2{\vcenter{\openup1\jot
  \halign{\strut #1\cr #2 \cr}}}

\def\fixit#1{}

\def\mop#1{\mathop{\rm #1}\nolimits}

\def\Tr{\mop{Tr}}

\def\href#1#2{#2}  

\def\lbldef#1#2{\expandafter\gdef\csname #1\endcsname {#2}}

\begin{document}
\baselineskip=16pt
\pagestyle{plain}
\setcounter{page}{1}

\begin{titlepage}

\begin{flushright}
hep-th/0210114 \\
PUPT-2053
\end{flushright}
\vfil

\begin{center}
{\huge AdS Dual of the Critical $O(N)$ Vector Model}
\end{center}

\vfil
\begin{center}
{\large  
I. R. Klebanov\footnote{E-mail: \tt klebanov@princeton.edu}
and
A. M. Polyakov\footnote{E-mail: \tt polyakov@princeton.edu}
}\end{center}

$$\seqalign{\span\TL\, & \sl\span\TT}{
 & Joseph Henry Laboratories, Princeton University, 
 Princeton, NJ  08544, USA  \cr
}$$
\vfil

\begin{center}
{\large Abstract}
\end{center}

We suggest a general relation between theories of infinite number
of higher-spin massless gauge fields in $AdS_{d+1}$ and large $N$
conformal theories in $d$ dimensions containing $N$-component vector
fields. In particular, we propose that the 
singlet sector
of the well-known critical 3-d $O(N)$ model with 
the $(\phi^a \phi^a)^2$ interaction is dual,
in the large $N$ limit, to the minimal bosonic
theory in $AdS_4$ containing massless gauge fields of even spin.

\vfil
\begin{flushleft}
October 2002
\end{flushleft}
\end{titlepage}
\newpage
\renewcommand{\thefootnote}{\arabic{footnote}}
\setcounter{footnote}{0}
\renewcommand{\baselinestretch}{1.2}  
\tableofcontents

\section{Introduction}
\label{Introduction}

It has long been anticipated that there exist exact dualities between
large $N$ field theories and strings \cite{thooft,Sasha}.
A gauge theory in $d$ dimensions is expected to be described by
a string background with $d+1$ non-compact curved dimensions \cite{Sasha}.
The particular cases of this duality that are best understood relate 
4-d conformal large $N$ gauge theories to type IIB strings on 
$AdS_5\times X_5$, where $X_5$ is a compact 5-d Einstein space 
\cite{juanAdS,gkPol,witHolOne}. For large `t Hooft coupling
$g_{\rm YM}^2 N$ many gauge theory 
observables may be calculated using the
supergravity approximation to this string theory.

For general `t Hooft coupling
this duality is still far from being understood completely.
One of the reasons is that it relates two very complicated 
theories. It is of some interest, therefore, to look
for simpler models or simpler limits realizing the AdS/CFT correspondence.
In this article we try to do just that. We consider the
large $N$ limit of the
$(\phi^a \phi^a)^2$ theory in 3-d space 
where $\phi^a$ is an $N$-component scalar field transforming
in the fundamental representation of $O(N)$.
It is well-known that this theory, which describes
critical points of $O(N)$ magnets, is conformal \cite{BW,WK}.
We conjecture that is has a dual $AdS_4$ description in terms
of a theory with infinite number of massless higher-spin
gauge fields.
Study of such theories has been going on for many years.
After the early work of Fronsdal \cite{Fron}, Fradkin and Vasiliev
\cite{FV}
formulated an interacting theory of infinitely many such fields in $AdS_4$.
Since then these theories were studied and generalized in a variety of
ways (for a review, see \cite{Vas}).
 
After the AdS/CFT correspondence was formulated, new
ideas emerged on the manifestation of the infinite number of conservation
laws that appear in the weakly coupled field theory 
\cite{Sundborg,EW,KVZ,SV,Sezgin,Misha,polWord,Mikh,Segal}.
Generally, it is expected that a conserved current
in the boundary theory corresponds to a massless gauge field in the
bulk \cite{ff}.
In particular, it was proposed that the massless higher-spin
gauge theory with ${\cal N}=8$ supersymmetry in $AdS_5$ is 
closely related to the free ${\cal N}=4$ large $N$ SYM theory
\cite{Sundborg,EW,Sezgin,Misha,Mikh}. This free theory
has an infinite number of conserved currents of increasing spin,
which are bilinears of the form\footnote{
This formula is schematic; the precise expression may be found, for instance,
in \cite{Mikh}.}     
\begin{equation}
J_{(\mu_1 \cdots
     \mu_s)}= \sum_{i=1}^6 \Tr  \Phi^i \nabla_{(\mu_1} \cdots 
     \nabla_{\mu_s)} \Phi^i + \ldots \,.
\end{equation}
where $\Phi^i$ are the six scalar fields in the adjoint representation
of $SU(N)$.
These currents 
are expected to be dual to the massless gauge fields in $AdS_5$
\cite{Sundborg,EW,Sezgin,Misha,Mikh}.

The correspondence between free CFT's of matrix-valued fields
and higher-spin massless gauge theories, suggested in
\cite{Sundborg,EW,KVZ,SV,Sezgin,Misha,Mikh}
and recently further developed in
\cite{Sundell}, is a remarkable conjecture, and we will
make use of a similar statement for free vector-valued fields in Section 2.
There is an essential difference, however, between the
adjoint and the fundamental {\it interacting} fields.  
In the adjoint case there is an exponentially growing
number of single-particle states in AdS corresponding to
single-trace operators of schematic form
\be
\Tr [\Phi \nabla^{l_1}\Phi \nabla^{l_2} \Phi \ldots \nabla^{l_k}
\Phi]
\ .
\ee
For any non-zero Yang-Mills coupling, 
operator products of
the currents bilinear in the adjoint fields
contain the whole zoo of such more complicated operators.
Theories of Fradkin-Vasiliev type do not contain enough
fields in AdS to account for all operators of this type. Hence, only
an appropriate generalization of the ${\cal N}=8$ supersymmetric
Fradkin-Vasiliev theory in $AdS_5$, with an infinite class of fields
added to it, may be dual to the weakly coupled 
${\cal N}=4$ large $N$ SYM theory.

In this paper we point out that theories of Fradkin-Vasiliev type
do contain enough fields to be dual to large $N$ field theories where
elementary fields are in the {\it fundamental} rather than 
adjoint representation. In this case, the only possible class of 
``single-trace'' operators are  $\phi^a \partial^l \phi^a$ 
whose number does not grow with the dimension (in contrast to the exponential
growth found for adjoint quanta). 
Effectively, there is only one ``Regge trajectory'' instead of
infinitely many. This roughly matches the number of 
fields found in theories of Fradkin-Vasiliev type.
Therefore, a massless higher-spin gauge theory in $AdS_{d+1}$
may capture the dynamics of such singlet currents.
A particularly simple picture of this duality
appears to emerge for the $AdS_4$ case which we address in the next section.

\section{$AdS_4$ and Vector Theories}

Little is known to date about the holographic duals
of massless higher-spin gauge theories in $AdS_4$ which are considerably
simpler than those in $AdS_5$. In this paper we propose that 
they are dual to large $N$ conformal field theories containing 
$N$-component vector rather
than $N\times N$ matrix fields. The simplest such $O(N)$
invariant theory is free:
\begin{equation}
S = {1\over 2}\int d^3 x \sum_{a=1}^N (\partial_\mu \phi^a)^2
\ .
\end{equation}
This theory has a class of $O(N)$ singlet conserved currents
\begin{equation}
\label{singlet}
J_{(\mu_1 \cdots \mu_s)}=   \phi^a \partial_{(\mu_1} \cdots 
     \partial_{\mu_s)} \phi^a + \ldots \,.
\end{equation}
There exists one conserved current for each even spin $s$.
We note that this spectrum of currents is in one-to-one correspondence
with the spectrum of massless higher-spin fields in the minimal
bosonic theory in $AdS_4$, the one governed by the algebra 
$ho(1;0|4)$ of \cite{MV} denoted $hs(4)$ in
\cite{ESS}. This theory, which contains
one massless gauge field for each even spin $s$,
is a truncation of the bosonic theory
containing one massless gauge field for each integer spin, governed by
the algebra $hu(1;0|4)$ \cite{MV,ESS}. 
The non-linear action for these fields is also known \cite{MV,ESS} but
it is not easy to extract explicit expressions from the
existing literature (the cubic interactions should be
constrained by the gauge invariance). 
We would like to conjecture that the correlation
functions of the {\it singlet} currents in the free 3-d theory may
be obtained from
the bulk action in $AdS_4$ through the usual AdS/CFT prescription
which identifies the boundary values of the fields with sources 
$h_0^{(\mu_1 \cdots \mu_s)}$ in
the dual field theory:
\begin{equation}
\langle \exp \int d^3 x\ 
h_0^{(\mu_1 \cdots \mu_s)} J_{(\mu_1 \cdots \mu_s)} \rangle =
e^{ S[h_0]}\ .
\end{equation}
$S[h_0]$ is the action of the high-spin gauge theory in $AdS_4$
evaluated
as a function of the boundary values of the fields.
This conjecture is closely related to similar suggestions in
\cite{Sundborg,EW,KVZ,SV,Sezgin,Misha,Mikh,Sundell}
on connections between theories of massless higher-spins in
$AdS_{d+1}$ and free fields in $d$ dimensions.

From the field theory point of view all correlators are given by
one-loop diagrams with the fields $\phi^a$ running around the loop,
so they may be evaluated exactly. 
Calculations are simple in position space where we may use the
propagator
\begin{equation}
\langle \phi^a (x_1) \phi^b (x_2)\rangle = {\delta^{ab}\over x_{12}}
\ ,
\end{equation}
where $x_{12} = |x_1 - x_2|$.
For example, for the correlators of the spin zero ``current''
$J = \phi^a \phi^a$ we then obtain
\begin{equation}
\langle J (x_1) J (x_2)\rangle \sim  {N\over x_{12}^2}
\ ,
\end{equation}
\begin{equation}
\langle J (x_1) J (x_2) J (x_3)\rangle 
\sim  {N\over x_{12} x_{13} x_{23} }
\ ,
\end{equation}
etc. The dimension of $J$ is $1$. The fact that it lies below
$d/2=3/2$ reveals a subtlety in building an AdS/CFT correspondence
for this field \cite{KWnew}: we have to use the negative branch of the 
formula for the dimension,
\be
\label{dimen}
\Delta_- = {d\over 2}-\sqrt{ {d^2\over 4} + (mL)^2 }\ ,
\ee
where $L$ is the radius of $AdS_4$.
To obtain $\Delta_-=1$ in $d=3$ we need $m^2=-2/L^2$. This corresponds
to a conformally coupled scalar field in $AdS_4$.
Perhaps this is the correct extension of the definition
of masslessness to spin zero.
Therefore, up to cubic order, we expect the following effective
Lagrangian for a scalar field $h$ in $AdS_4$ dual to the scalar
current $J$:
\begin{equation}
S= {N\over 2}
\int d^4 x \sqrt g \left [ (\partial_\mu h)^2 + {1\over 6} R h^2
+ \alpha h^3 + \ldots   \right]\ .
\end{equation}
Since $R=-12/L^2$ in $AdS_4$, we indeed find $m^2=-2/L^2$.

As explained in \cite{KWnew}, the unconventional branch
$\Delta_-$ introduces a subtlety into the procedure for extracting
the correlation functions. The correct procedure is to first work out
the generating functional $W[h_0, \ldots]$ for
correlation functions with the conventional dimension
$\Delta_+$ for the operator dual to $h$, and then to carry out
the Legendre transform with respect to the 
source $h_0$ \cite{KWnew}.

This begs the question: what is the physical meaning of the theory
where the operator $J$ has the conventional dimension
$\Delta_+=2$? The answer turns out to be interesting: this CFT
is the well-known fixed point of the 
{\it interacting} $O(N)$ vector
model with the 3-dimensional action
\begin{equation}
\label{son}
S = \int d^3 x \left [{1\over 2} (\partial_\mu \phi^a)^2 + 
{\lambda\over 2N} (\phi^a \phi^a)^2  \right ]
\ .
\end{equation}
 The standard trick for dealing with this interaction is
to introduce an auxiliary field $\sigma(x)$ so that the action
assumes the form
\begin{equation}
S = \int d^3 x \left [{1\over 2} (\partial_\mu \phi^a)^2 +
\sigma \phi^a \phi^a - {N\over 2\lambda } \sigma^2 \right ]\ .
\end{equation}
Now the action is quadratic in 
the fields $\phi^a$ and integrating over them one finds the
effective action for $\sigma$. This provides an efficient way
of developing the $1/N$ expansion \cite{WK}.

Note that the interaction term may be written as
$\lambda J^2/(2N)$. 
 This is a vector model analogue of the
trace-squared terms which have been recently studied in the AdS/CFT
setting \cite{ABS,Ed,BSS,Steve}. In \cite{Ed} it was shown, 
using boundary conditions in AdS, that when a {\it relevant} interaction
of this kind is added to the action, then the theory flows
from an unstable UV fixed point where $J$ has dimension $\Delta_-$
to an IR fixed point where $J$ has dimension 
$\Delta_+$.\footnote{There is an analogous phenomenon in
2-d Liouville gravity: change of the branch of gravitational dressing
caused by adding a trace-squared operator to the matrix model action
\cite{me}.} In \cite{Steve} this 
type of flow was studied in more detail
and further evidence for it was provided.
The interaction is relevant in the UV because the dimension of operator
$J^2$ is $2 \Delta_- + O(1/N)$, and from (\ref{dimen})
it is clear that $2\Delta_- < d$. Similarly, it is clear that the
interaction is irrelevant in the IR where the dimension 
of the operator $J^2$ is $2 \Delta_+ + O(1/N)$, and
we generally have $ 2\Delta_+ > d$. For this
reason the presence of this operator does not
produce a line of fixed points.

The flow from a free $O(N)$ vector model to the interacting
model (\ref{son}) is an example of the general discussion above. 
In fact, it has been known for many years that,
at the IR critical point, the operator $J$ has dimension 
$\Delta_J = 2+O(1/N)$ 
\cite{WK}.
For large $N$ this value coincides with $\Delta_+$ that is required
by the AdS analysis of \cite{Ed}.
Furthermore, this isolated IR fixed point exists not only in the large $N$
limit, but also for any finite $N$. So, in this case
the RG flow produced by the addition of operator $J^2$ is not destabilized
by $1/N$ corrections. 

Therefore, we make the following conjecture. Suppose that we
start with an action in $AdS_4$ that describes the 
minimal bosonic higher-spin gauge theory with 
even spins only and symmetry group $ho(1;0|4)=hs(4)$.
If we apply the standard AdS/CFT methods to this action, using
dimension $\Delta_+$ for all fields, then we find the correlation
functions of the singlet currents in the interacting large $N$
vector model (\ref{son}) at its IR critical point. 
A weak test of this conjecture is that the anomalous
dimensions of all the currents with spin $s>0$ are of order $1/N$ \cite{WK}
(for the stress-energy tensor, $s=2$, it is exactly zero),
so in the large $N$ limit they correspond to massless
gauge fields in the bulk. For example, all planar 3-point functions
of currents with $s>0$ are exactly the same in the interacting 
$O(N)$ theory as in the free theory. This shows why in the interacting
theory all these currents are conserved to leading order in $N$.

Another argument in favor of our conjecture is the following.
If we Legendre transform the generating functional of the interacting
large $N$ vector model with respect to 
the source $h_0$ that couples to $J$, then we obtain the
generating functional of singlet current correlators in the 
{\it free} vector theory. On the AdS side of this duality this statement
follows from the rule worked in \cite{KWnew} for operators with
dimension $\Delta_-$. On the field theory side, the Legendre transform
removes the diagrams one-particle reducible
with respect to the auxiliary field $\sigma$, so
that only the free field contributions to the singlet correlators
remain.

\section{Operator Products at Large $N$}

The operator structure at large $N$ has some unusual
features. First of all, we expect that operators come in two types,
elementary and composite. In the case of gauge theory
they roughly correspond to the single-trace and
multi-trace operators (we say ``roughly'' because in general
single- and multi-trace operators mix).
In the dual AdS language they correspond to one-particle and
multi-particle states.

Let us first clarify why these structures are inevitable at 
large $N$.\footnote{This part of the discussion is closely related
to a similar discussion in \cite{Frolov}.}
Consider a set $\{ \Omega_k\}$
of single-trace operators in gauge theory or of
singlet bilinears (\ref{singlet}) in a vector theory.
Let us suppose for a moment that the algebra of such operators
closes. Then the standard large $N$ counting would imply that,
in the gauge theory,
\be\label{opgauge}
\Omega_k (x_1) \Omega_l (x_2) 
\sim N^2 \delta_{kl} x_{12}^{-2\Delta_k} I + f_{klm}
x_{12}^{\Delta_m -\Delta_k -\Delta_l} \Omega_m
\ ,
\ee
while in the vector theory $N^2$ is replaced by $N$.
However, these operator products are clearly inconsistent
with contributions of disconnected terms.
For example, in the 4-point function we have
\be
\langle \Omega_k \Omega_l \Omega_m \Omega_n \rangle
= 
\langle \Omega_k \Omega_l \rangle\langle \Omega_m \Omega_n \rangle +
\langle \Omega_k \Omega_m \rangle\langle \Omega_l \Omega_n \rangle +
\langle \Omega_k \Omega_n \rangle\langle \Omega_l \Omega_m \rangle  +
\langle \Omega_k \Omega_l \Omega_m \Omega_n \rangle_{\rm conn}
\ee
In the gauge theory the disconnected terms are of order $N^4$
while the connected ones are of order $N^2$;
in the vector theory
the disconnected terms are of order $N^2$
while the connected ones are of order $N$.
As we substitute the OPE (\ref{opgauge})
into the left-hand side of the 4-point function, the
unit operator will reproduce the first disconnected
term and the remaining $\Omega_l$ will contribute to
the connected term. However, the two remaining disconnected
terms representing the unit operators in the cross channels
remain unaccounted for!

This forces us to add composite 
``double-trace'' operators $\Omega_{kl}$
on the right-hand side of the operators products
(\ref{opgauge}). Repeating this argument
for higher-point functions, we will see the need for all composite
``multi-trace'' operators $\Omega_{k_1 \cdots k_n}$.
Their correlation functions are defined so as to reproduce the 
disconnected contributions to correlators.
The crucial difference between the elementary and the composite
operators is that the dimensions of the latter
are determined by the dimensions of the former, up to
$1/N$ corrections:
\be
\Delta (\Omega_{kl}) = \Delta_k + \Delta_l + {1\over N^2} \omega_{kl} +
\ldots ,
\ee
etc. In the vector model, $1/N^2$ is replaced by $1/N$.

Let us briefly describe the structure of the operator
algebra in the interacting $O(N)$ vector model.
First of all, as we already mentioned,
the operator $J=\phi^a \phi^a$ has dimension
$\Delta_J= 2+ O(1/N)$, while in the free theory its dimension would be
$1$. It is not hard to check that the composite operators
$J^p$ have dimensions
\be
\Delta_p = p \Delta_J + O(1/N)
\ee
in accordance with the general arguments above.
In the large $N$ limit $\langle J\rangle$ is proportional to the 
auxiliary
 field $\sigma$ used to solve the model \cite{WK}.
Let us consider the 4-point function 
\be \label{fourpt}
\langle J(x_1) J(x_2) J(x_3) J(x_4) \rangle 
\ .
\ee
We first recall that for any 3 conformal primary operators we have
the formula
\be
\langle A(x_1) B(x_2) C(x_3) \rangle
\sim {f_{ABC}\over x_{12}^{\Delta_A + \Delta_B - \Delta_C}
x_{13}^{\Delta_A + \Delta_C - \Delta_B}
x_{23}^{\Delta_B + \Delta_C - \Delta_A}
}
\ .
\ee
The contribution of operator $O$ into the 4-point function 
of $A$, $B$, $C$ and $D$ is given
by
\be
\int d^d x  \langle A(x_1) B(x_2) O(x) \rangle
\langle \bar O(x) C(x_3) D(x_4) \rangle
\ .\ee
Here $O(x)$ is an operator of dimension $\Delta$ while
$\bar O(x)$ is its conjugate of dimension $\bar\Delta=d-\Delta$.
If we take the limit $x_{12}, x_{34} \rightarrow 0$
to uncover the OPE, we find from this formula
\be \label{unwanted}
\langle A(x_1) B(x_2)
 C(x_3) D(x_4) \rangle \sim
 {1\over \Delta -\bar \Delta} x_{12}^{-\Delta_A-\Delta_B}
 x_{34}^{-\Delta_C-\Delta_D}
 \left \{ \bigg ({x_{12} x_{34}\over x_{13}^2} \bigg )^\Delta -
\bigg ({x_{12} x_{34}\over x_{13}^2}\bigg )^{\bar\Delta} \right \}
\ .
\ee
The second term is an unwanted contribution of an operator
with dimension $\bar\Delta$ to the OPE. 
The presence of such ``shadow'' contributions was noted in
the context of the $O(N)$ model in \cite{SP} and later on analyzed
in the large $N$ limit in \cite{Ruhl,Petkou}.
In \cite{SP} it was shown using dispersion relations
that one can construct a conformal amplitude different from
(\ref{unwanted}). It does not contain the contribution
of dimension $\bar\Delta$ but contains terms $\sim
\log (x_{12}^2 x_{34}^2)$ which originate from the
contribution of composite operators. In \cite{SP}
the requirement of cancellation of the logarithmic terms 
between the connected and the disconnected contributions
to the correlator gave the
bootstrap condition determining anomalous dimensions and structure
constants.
In AdS calculations of 4-point functions 
of 1-particle operators, the logarithmic terms
of the type described above were found in \cite{Freed,FH}.
It is not hard to check that the ``unitary amplitude''
of \cite{SP} and the AdS amplitude have the same form.

In the case of the 
interacting $O(N)$ vector model these general considerations
simplify greatly. 
The 4-point function (\ref{fourpt}) is given by the sum of
disconnected pieces, 3 
exchange diagrams with an intermediate
auxiliary field line, and the box diagram corresponding
to the loop of the field $\phi^a$.  
The dimension of $\phi^a$ at the IR critical point
is $\Delta_\phi = 1/2+
O(1/N)$ \cite{WK}. As we take the limit
$x_{12}, x_{34} \rightarrow 0$, we find that
the box diagram behaves as
$ (x_{12}x_{34})^{2\Delta_\phi - 2\Delta_J} x_{13}^{-4\Delta_\phi} $. Also,
the exchange diagram has an unwanted contribution
$ (x_{12}x_{34})^{\bar\Delta - 2\Delta_J} x_{13}^{-2\bar \Delta} $,
where $\bar \Delta = d - \Delta_J = 1 + O(1/N)$.
The correct OPE structure is possible only if these terms cancel
each other. Hence, in the large $N$ limit we must have
$\bar\Delta = 2 \Delta_\phi$, which is indeed the case!

Most importantly, the contributions of the higher-spin currents
$J_{(\mu_1 \cdots \mu_s)}$ with $s>0$
appear from the higher-order terms in the expansion of
the box diagram in $x_{12}$ and $x_{34}$.
In this way we see explicitly how the presence of the
infinite number of higher-spin fields in $AdS_4$ is necessary to
reproduce the OPE of the critical $O(N)$ vector model.
It remains to be seen whether they
are precisely related to the $AdS_4$ theory of \cite{MV,ESS}
via the AdS/CFT correspondence.

Finally, we indicate how the discussion of the 4-point function
is modified if we simply consider the free theory of the 
scalar fields $\phi^a$.
Then there are no diagrams where the auxiliary field
is exchanged, hence no unwanted terms of the form
$ (x_{12}x_{34})^{\bar\Delta - 2\Delta_J}
x_{13}^{-2\bar \Delta} $ that need to be
canceled. The leading term in the box diagram, which
is now exactly given by $(x_{12} x_{23} x_{34} x_{41} )^{-1}$,
is still of the form
$ (x_{12}x_{34})^{2\Delta_\phi - 2\Delta_J}
x_{13}^{-4\Delta_\phi} $, but now
$\Delta_J = 2 \Delta_\phi = 1$. Hence this term correctly
reproduces the contribution of operator $J$ to the OPE.
The subleading terms in the expansion of the box diagram
correspond to the contribution of the currents 
$J_{(\mu_1 \cdots \mu_s)}$ with $s>0$.

\section{Discussion}

There is a number of possible extensions of the duality proposed
above. It is not hard to construct a $U(N)$
invariant theory which contains
one singlet current for each integer spin $s$.
This theory has one
{\it complex} $N$-component field $\varphi^a$.
Then, in addition to the currents of even spin,
\begin{equation}
J_{(\mu_1 \cdots
     \mu_s)}= \varphi^{*a}\partial_{(\mu_1} \cdots 
     \partial_{\mu_s)} \varphi^a + \ldots + c.c. ,
\end{equation}
we find currents of odd spin
\begin{equation}
J_{(\mu_1 \cdots
     \mu_s)}= 
     i\varphi^{*a}\partial_{(\mu_1} \cdots 
     \partial_{\mu_s)} \varphi^a + \ldots + c.c. .
\end{equation}
We expect this theory to be dual to the bosonic $AdS_4$ theory constructed 
in \cite{MV}, without the projection that throws away the odd spins.
In the classification of \cite{MV}
this non-minimal bosonic higher-spin algebra 
is $hu(1;0|4)$ while in \cite{ESS} it was denoted $hs_0(4;1)$.

A theory corresponding to the extended higher-spin algebra
$hu(n;0|4)$ \cite{MV}
may be constructed out of $n$ complex $N$-component
fields $\varphi^a_I$, $I=1, \ldots, n$. Such a theory possesses
a set of spin $1$ currents in the adjoint representation of $U(n)$,
\begin{equation}
i\varphi^{*a}_I\partial_\mu \varphi^a_J - i (\partial_\mu
\varphi^{*a}_I) \varphi^a_J \ .
\end{equation}
Therefore, the corresponding bulk $AdS_4$ theory has $U(n)$
gauge symmetry. Similarly, higher-spin gauge theory based on
the algebra $ho(n;0|4)$ \cite{MV} has $O(n)$ gauge symmetry in $AdS_4$
and should be dual to $O(N)$ field theory with $n$
real $N$-component scalar fields in $d=3$.

If we supplement the field theory with fermionic 
$N$-component fields,
then we also find currents of half-integral spin.
Since supersymmetric higher-spin theories in $AdS_4$ are
well-known \cite{Vas,KVZ,ESS}, it would be interesting to work out the 
supersymmetric 
analogues of the duality in detail.

From the large $N$ field theory point of view, there is an obvious 
generalization from $d=3$ to $d=4-\epsilon$.
The critical points of vector theories are well-known to exist
for $0<\epsilon < 2$. Perhaps there is a sense in which these theories are dual
to bulk theories in $5-\epsilon$ dimensional AdS space.

\section*{Acknowledgments}
We thank L. Rastelli and E. Witten for useful discussions. 
I.R.K. is also grateful to the Institute for Advanced Study
for hospitality.
This material is based upon work
supported by the National Science Foundation Grant No.
PHY-9802484.  
Any opinions, findings, and conclusions or recommendations expressed in
this material are those of the authors and do not necessarily reflect
the views of the National Science Foundation.

\begingroup\raggedright\endgroup


\end{document}